\documentclass[journal=jacsat,manuscript=article]{achemso}

\usepackage{chemformula} 
\usepackage[T1]{fontenc} 
\usepackage{graphicx}
\usepackage{soul} 

\graphicspath{
{./figure/}
{./}
}

\author{Kairi Masuda}
\email{kairi.masuda.c5@tohoku.ac.jp}
\author{Yu Kumagai}

\affiliation[Tohoku University]
{Institute for Materials Research, Tohoku University, 2-1-1 Katahira, Aoba-ku, Sendai, 980-8577, Japan}

\title[An \textsf{achemso} demo]
{Atomic-scale phase-field modeling\\with universal machine learning potentials}

\abbreviations{IR,NMR,UV}
\keywords{American Chemical Society, \LaTeX}

\begin{document}







\begin{abstract}
Atomic-scale phase-field modeling formulates the probability densities of atomic vibrations as Gaussian distributions and derives a free energy functional using variational Gaussian theory and interatomic potentials. This framework permits per‑Gaussian decomposition of the free energy, providing a description of local thermodynamic states with atomic resolution. However, existing formulations are limited to classical pairwise interatomic potentials, restricting their applicability to specific materials and compromising quantitative accuracy. In this work, we extend the atomic-scale phase-field methodology by incorporating universal machine learning interatomic potentials, thereby generalizing the free energy functional to many-body systems. This extension enhances both the accuracy and transferability of the approach. We demonstrate the method by applying it to bulk copper under NVT and NPT ensembles, where the predicted pressures and equilibrium lattice constants show excellent agreement with molecular dynamics simulations, validating the theoretical framework. Furthermore, we apply the method to $\Sigma$5(310)[001] grain boundaries in copper, enabling the visualization of local free energy distributions with atomic-scale resolution. The results reveal a pronounced free energy concentration at the grain boundary core, capturing the thermodynamic signature of the interface. This study establishes a versatile and accurate framework for atomic-scale thermodynamic modeling, significantly broadening the scope of phase-field approaches to include complex materials and defect structures.
\end{abstract}


\clearpage

\section{Introduction}
In phase-field modeling, free energies, which are typically calculated as a global property in other methodologies such as molecular dynamics\mbox{\cite{FrenkelLadd,BENNETT1976245,BONOMI20091961}}, are formulated as position-dependent functional. The system then evolves to minimize this spatially inhomogeneous free-energy landscape at each position. This local thermodynamic framework has proved highly effective for simulating microstructural evolution such as solidification, fracture, domain formations in ferroelectrics, and so on\mbox{\cite{LQchenReview,SolidificationReview,LQchen_review_thin_film,Ambati2015,doi:10.1098/rsta.2015.0166}}. Phase-field simulations not only reproduce such behavior but also supply spatially resolved thermodynamic information, including local free energy, elastic energy, and electrostatic energy, that are difficult to obtain experimentally\mbox{\cite{CHOUDHURY20055313,Gao2013,Park2018,WANG2021117383,NatureYang2025,doi:10.1126/science.abb3209}}. As a result, phase-field modeling serves as a computational microscopy for local thermodynamic states in materials. On the other hand, recent breakthroughs in materials science, such as the discovery of 2D materials, advances in high-resolution microscopy, the emergence of high-entropy alloys, and renewed interest in molecular ferroelectrics have shifted attention toward smaller length scales where heterogeneity extends down to individual atoms\mbox{\cite{2D_ferroelectric,STEM_ferroelectrics,PTO_dislocation,MIRACLE2017448,Horiuchi2008,elastin}}. Conventional phase-field modeling, rooted in continuum mechanics, cannot be directly applied to such atomistic systems. This suggests that a local thermodynamic framework with true atomic resolution is essential for further progress.

Atomic-scale phase-field modeling addresses this need by representing the probability density of atomic vibrations as Gaussian distributions and constracting the free energy functional based on variational Gaussian theory and interatomic potentials\mbox{\cite{PhysRevLett.63.624,doi:10.1063/1.460547,PhysRevB.84.054103,PhysRevB.110.104107}}. This formulation enables the free energy to be decomposed into contributions from individual Gaussians, thereby providing atomic-level insight into local thermodynamic states. Previous implementations, however, have relied on classical pairwise potentials, limiting both accuracy and transferability. Concurrently, the high-throughput density-functional databases are rapidlly grown\mbox{\cite{10.1063/1.4812323,SCHMIDT2024101560,doi:10.1021/acscatal.2c05426}} and this has enabled “universal” machine-learning interatomic potentials that predict energies and forces across broad chemistries with first-principles accuracy at modest calculation cost\mbox{\cite{Zeni2025,Batatia2025,Deng2023,Takamoto2022}}. Incorporating such universal potentials into atomic-scale phase-field modeling would transform it into a general, high-fidelity local thermodynamic theory applicable to virtually any material system.

Here, we have developed atomic-scale phase-field modeling incorporating universal machine learning interatomic potentials, aiming to achieve both broad applicability and high accuracy. This paper is organized as follows: First, we extend the free energy functional to many-body systems, enabling the use of universal machine learning potentials, and describe computational details. Next, we apply the proposed methodology to bulk copper (Cu) under NVT and NPT ensembles, comparing predicted pressures and lattice constants with molecular-dynamics simulations to validate the theory. We then describe the procedure for decomposing the free energy and apply the method to $\Sigma$5(310)[001] grain boundaries in copper to visualize local thermodynamic states with atomic resolution. Finally, we summarize the key findings and discuss future prospects of this approach.

\clearpage
\section{Methods}

\subsection{Theory}
The thermal vibration of an atom around a mean position under harmonic thermal motion appears as a density cloud on a longer time scale. These probability clouds make Gaussian distributions\mbox{\cite{PhysRevLett.63.624,doi:10.1063/1.460547,PhysRevB.84.054103,PhysRevB.110.104107}}, of which the general form can be described as follows:
\begin{eqnarray}
\rho_{i}(\boldsymbol{x}_{i}|\boldsymbol{X}_{i}, \boldsymbol{\Sigma}_i)&=&\frac{1}{(2\pi)^{3/2}\sqrt{|\boldsymbol{\Sigma}_{i}|}}\\ \nonumber
&\times&\exp  \Big\lbrace -\frac{1}{2}(\boldsymbol{x}_{i}-\boldsymbol{X}_{i})^T \boldsymbol{\Sigma}^{-1}_{i}(\boldsymbol{x}_{i}-\boldsymbol{X}_{i}) \Big\rbrace 
\end{eqnarray}
where $\boldsymbol{X}_{i}=(X_{i},Y_{i},Z_{i})$ is the position of the vibration center of atoms $i$ =1, 2, ..., $N$. $\boldsymbol{\Sigma}_{i}$ is a covariance matrix as follows:
\begin{eqnarray}
\boldsymbol{\Sigma}_{i}=\begin{pmatrix}
\sigma_{i,x}^{2} & \rho_{i,xy}\sigma_{i,x}\sigma_{i,y}  & \rho_{i,xz}\sigma_{i,x}\sigma_{i,z} \\
 \rho_{i,xy}\sigma_{i,x}\sigma_{i,y} & \sigma_{i,y}^{2}  &  \rho_{i,yz}\sigma_{i,y}\sigma_{i,z}  \\
 \rho_{i,xz}\sigma_{i,x}\sigma_{i,z} & \rho_{i,yz}\sigma_{i,y}\sigma_{i,z} & \sigma_{i,z}^{2}
\end{pmatrix}
\end{eqnarray}
where $\sigma_{i,x}$, $\sigma_{i,y}$, $\sigma_{i,z}$ are the standard deviations along $x$, $y$, and $z$ directions while $\rho_{i,xy}$, $\rho_{i,yz}$, and $\rho_{i,xz}$ are correlation coefficients which represent the Gaussian spatial orientation. $|\boldsymbol{\Sigma}_{i}|$ is the determinant, and $\boldsymbol{\Sigma}_{i}^{-1}$ is the inverse of the covariance matrix. When $\boldsymbol{X}_{i}$ and $\boldsymbol{\Sigma}_i$ are given, a probability density at the position $\boldsymbol{x}_{i}$ is produced.



The variational Gaussian theory by LeSar \textit{et al.}, which calculates the free energies of solids based on the Gibbs-Bogoliubov inequality, was originally formulated for isotropic Gaussians\mbox{\cite{PhysRevLett.63.624,doi:10.1063/1.460547,PhysRevB.84.054103}} using a Morse potential for Cu, that is, a classical pair potential. Here, we extend the variational Gaussian theory for the general Gaussians\mbox{\cite{PhysRevB.110.104107}} and combine it with many-body systems for a universal machine learning potential to derive the free energy of any materials with high-accuracy. That is, the free energy can be analytically represented as follows:
\begin{eqnarray}
\label{eq:freeene}
F&=&\frac{1}{2}k_{B}T\sum_{i=1}^{N} \left \{ \ln \left(\frac{\Lambda_{i}^{6}}{8\pi^{3} \sigma_{i,x'}^2 \sigma_{i,y'}^2 \sigma_{i,z'}^2} \right)-3 \right\} \\
&+&\int\int...\int{\rho_{1}(\boldsymbol{x}_{1})}{\rho_{2}(\boldsymbol{x}_{2})}...{\rho_{N}(\boldsymbol{x}_{N})}\phi(\boldsymbol{x}_{1},\boldsymbol{x}_{2},...,\boldsymbol{x}_{N})d\boldsymbol{x}_{1}d\boldsymbol{x}_{2}...d\boldsymbol{x}_{N}. \nonumber
\end{eqnarray}
The first term is the energy for atomic vibrations, where $\Lambda=\hbar\sqrt{2\pi/m_{i}k_{B}T}$ is the thermal de Broglie wavelength, that is, $\hbar$ is the Planck constant,$m_{i}$ is atomic mass of atoms $i$, and $k_{B}$ is the Boltzmann constant, and $T$ is temperature. $\sigma_{i,x'}^2$, $\sigma_{i,y'}^2$, and $\sigma_{i,z'}^2$ are dispersions along the principal axes, that is, the eigenvalues of $\boldsymbol{\Sigma}_{i}$. The second term is expectation value of multi-body potantial energy  $\phi$, that is, an effective potential between multi-Gaussians.

Under given positions $\boldsymbol{X}_{i}$, optimal Gaussian shapes $\boldsymbol{\Sigma}_{i}$ are calculated by minimizing the above free energy. Then, positions should move in time to reduce free energy while conserving the atomic probability density, that is, $\int \rho d\boldsymbol{x} = 1$. These imply that the probability density should move according to a conservative phase-field equation as follows:
\begin{eqnarray}
\label{eq:govern}
  \frac{\partial \rho_{i}(\textbf{\textit{x}},t)}{\partial t}&=&\nabla \cdot \left(D_{i}(\textbf{\textit{x}})\nabla \frac{\delta F}{\delta \rho_{i}} \right) \\ \nonumber
  &=& \nabla \cdot (\kappa_{i} \rho_{i}(\textbf{\textit{x}}) \nabla \Phi_{i}(\textbf{\textit{x}})),
\end{eqnarray}
where $t$ is time and $D_{i}(\textbf{\textit{x}})$ is a position-dependent kinetic coefficient. Here, as $\kappa_{i}$ is a kinetic coefficient, we set $D_{i}(\textbf{\textit{x}}) = \kappa_{i} \rho_{i}(\textbf{\textit{x}})$ because a probability density should follow the continuity equation. Here, a differential of the functional derivative $\nabla\delta F/\delta \rho_{i}$ = $\nabla \Phi_{i}(\textbf{\textit{x}})$ is an effective force. By solving the above governing equation, the motion of probability densities can be calculated.

\clearpage

\subsection{Simulation detail}
To validate the proposed theory, we selected bulk Cu as a model system and evaluated its pressure and lattice constant in both the canonical (NVT) and isothermal–isobaric (NPT) ensembles. A conventional-cell-based 2 $\times$ 2 $\times$ 2 face-centered-cubic supercell containing 32 atoms was constructed, and the atomic trajectories were calculated under each ensemble condition. The following section details the practical implementation of the theory in these simulations.



Figure 1 presents a schematic overview of the computational workflow. The second term in the free energy functional in Eq. (3) is a high-dimensional integral that is most efficiently evaluated by Monte Carlo techniques. Because the underlying probability density is a product of mutually independent Gaussian functions, we employ importance sampling: random configurations are drawn directly from the relevant Gaussian distributions rather than from a uniform distribution. Under this transformation the integral can be rewritten as follows\mbox{\cite{PhysRevB.89.064302,PhysRevB.98.024106}}:
\begin{eqnarray}
\label{eq:MonteCarloIntegral}
&&\int\int...\int{\rho_{1}(\boldsymbol{x}_{1})}{\rho_{2}(\boldsymbol{x}_{2})}...{\rho_{N}(\boldsymbol{x}_{N})}\phi(\boldsymbol{x}_{1},\boldsymbol{x}_{2},...,\boldsymbol{x}_{N})d\boldsymbol{x}_{1}d\boldsymbol{x}_{2}...d\boldsymbol{x}_{N} \\ \nonumber
&=&\frac{1}{N_{sample}}\sum_{i=1}^{N_{sample}} \frac{{\rho_{1}(\boldsymbol{x}_{1,i})}{\rho_{2}(\boldsymbol{x}_{2,i})}...{\rho_{N}(\boldsymbol{x}_{N,i})}\phi(\boldsymbol{x}_{1,i},\boldsymbol{x}_{2,i},...,\boldsymbol{x}_{N,i})}{{\rho_{1}(\boldsymbol{x}_{1,i})}{\rho_{2}(\boldsymbol{x}_{2,i})}...{\rho_{N}(\boldsymbol{x}_{N,i})}}\\ \nonumber
&=&\frac{1}{N_{sample}}\sum_{i=1}^{N_{sample}}\phi(\boldsymbol{x}_{1,i},\boldsymbol{x}_{2,i},...,\boldsymbol{x}_{N,i}),
\end{eqnarray}
where $\boldsymbol{x}_{1,i}, \boldsymbol{x}_{2,i},...,\boldsymbol{x}_{N,i}$ denotes the $i$-th configuration, whose atomic coordinates are independetly drawn from the Gaussian probability densities $\rho_{1}$, $\rho_{2}$, ..., $\rho_{N}$. To reduce the statistical error of the Monte-Carlo estimate, we employ a Sobol low-discrepancy sequence\mbox{\cite{SOBOL196786}} and map it to the Gaussian ensemble through the inverse cumulative distribution function, $N_{sample}= 1000$ in total. The interatomic potential $\phi$ is provided by the Multi Atomic Cluster Expansion (MACE) model\mbox{\cite{Batatia2025}}, which encodes each atomic configuration as a graph. Because MACE can evaluate many graphs concurrently, the integrand is computed in a parallel fashion on a GPU via PyTorch\mbox{\cite{paszke2017automatic}}, allowing the high-dimensional integral to be evaluated with minimal wall-time overhead. The gradient with respect to the Gaussian shape parameters $\Sigma$ is evaluated by the reparameterization trick (see Supplementary Information 1). Subsequent minimization of the free-energy functional is carried out with the L-BFGS algorithm implemented in SciPy\mbox{\cite{LBFGS,2020SciPy-NMeth}}, using a relative convergence tolerance of 1.0 $\times$ 10$^{-6}$.


Our multibody free-energy framework and the corresponding numerical procedure share conceptual similarities with the stochastic self-consistent harmonic approximation (SSCHA), which also employs importance sampling within a variational formulation to evaluate phonon free energies\mbox{\cite{PhysRevB.89.064302,PhysRevB.98.024106}}. A key difference lies in the choice of variational parameters. In the present work we optimize the Gaussian width tensors directly, introducing 6N independent variables for a system of N atoms, whereas SSCHA determines the full dynamical matrix, which involves 3N $\times$ 3N parameters. The substantially reduced parameter space renders our method more amenable to large-scale simulations and facilitates the explicit treatment of defects and other local perturbations.


For the phase-field equation, Eq. (\mbox{\ref{eq:govern}}) has the same form as the Fokker-Planck equation described as follows:
\begin{equation}
  \frac{\partial \rho_{i}(\boldsymbol{x},t)}{\partial t}=\nabla ( \boldsymbol{a}_{i}(\boldsymbol{x}) \rho_{i}(\boldsymbol{x})),
\end{equation}
where $\boldsymbol{a}$ is a drift coefficient. To make this equation simple, we assume that each Gaussian distribution moves without deformation during the infinitesimal motion, and then we update the Gaussian shape. Under such assumption, the equation can be converted into the following Langevin equation\mbox{\cite{PhysRev.36.823,PhysRevB.110.104107}}:
\begin{equation}
\label{eq:langevin}
  \frac{d\boldsymbol{X_{i}}}{dt}=-\kappa_{i} \nabla_{i} F(\boldsymbol{X_{i}}).
\end{equation}
We calculate the motion of the probability densities by solving this equation by the Verlet algorithm with the time step $\Delta t$ = 100 fs. The gradient of the free energy $\nabla F$ can be represented as follows:
\begin{eqnarray}
 -\nabla_{i} F &=&-\frac{\partial F}{\partial X_{i}} \\ \nonumber
 &=&-\int\int...\int{\rho_{1}(\boldsymbol{x}_{1})}{\rho_{2}(\boldsymbol{x}_{2})}...{\rho_{N}(\boldsymbol{x}_{N})}\frac{\partial \phi}{\partial X_{i}}(\boldsymbol{x}_{1},\boldsymbol{x}_{2},...,\boldsymbol{x}_{N})d\boldsymbol{x}_{1}d\boldsymbol{x}_{2}...d\boldsymbol{x}_{N} \\ \nonumber
 &=&\Big \langle-\frac{\partial \phi}{\partial X_{i
 }}\Big \rangle_{\rho_{1},\rho_{2},...,\rho_{N}},
\end{eqnarray}
Thus, the gradient of free energy, that is, mean force is an expecation value of force by interatomic potential. For mobility $\kappa$, we employed $\kappa=5.56\times10^{-8}$ m$^{2}$/(V$\cdot$ s), which is the mobility of Cu$^{2+}$ in water at 298 K\mbox{\cite{Cumobility,Ion_mobility_equation}}.
 
For comparison, we conducted the ordinal molecular dynamics (MD) simulation using the Atomic Simulation Environment (ASE) to calculate the time evolution of the above initial structure under NVT and NPT ensembles\mbox{\cite{ase-paper,ISI:000175131400009}} of 10000 steps with the time step being 1 fs. Values such as pressure are calculated at each time step.

 
 \begin{figure}
  \includegraphics{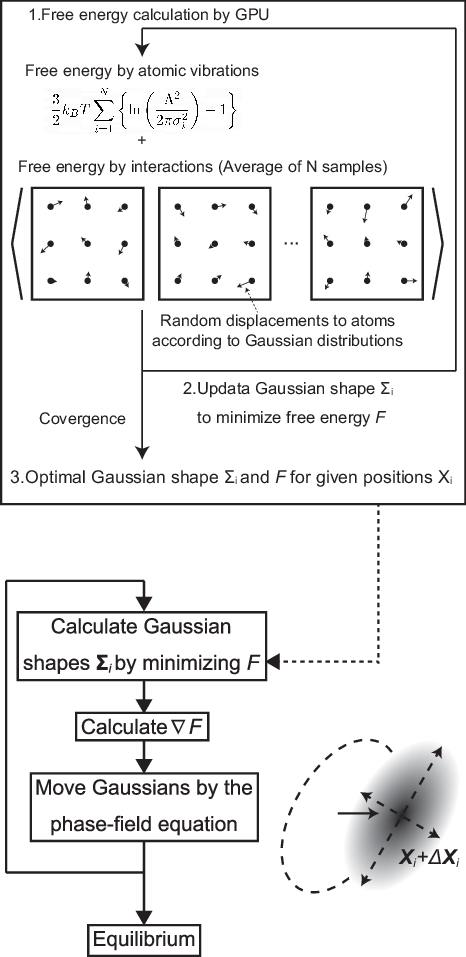}
  \caption{A schematic illustration of the computational workflow for phase-field modeling that includes many-body interactions.}
  \label{fig:nvttime}
\end{figure}

\clearpage

\section{Results}
\subsection{Time evolution}
\begin{figure}
  \includegraphics{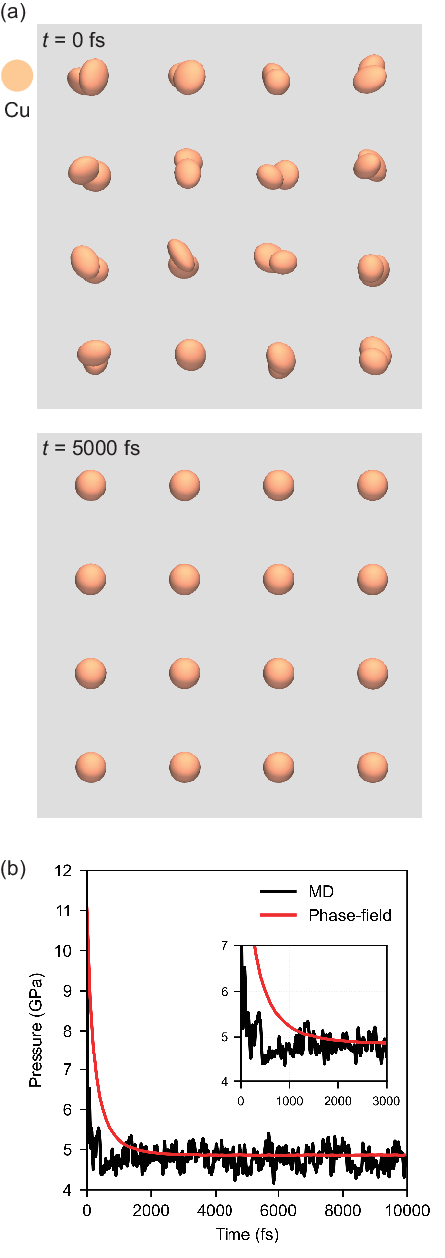}
  \caption{(a) Time evolution of the atomic probability density dispersion $\boldsymbol{\Sigma}_{i}$ for Cu atoms obtained by phase-field simulations under NVT conditions with the lattice constant of 3.615 \mbox{\AA} and $T$ = 300 K. Note that the dispersion of the atomic density is magnified fivefold. (b) Time evolution of the pressure obtained from the phase-field and MD simulations.}
  \label{fig:time}
\end{figure}

Figure \mbox{\ref{fig:time}}(a) and the Supplemental Material show the time evolution of the atomic densities in Cu under the NVT ensemble at temperature $T$ = 300 K and the experimental lattice constant of 3.615 \mbox{\AA} using phase-field simulations. Note that the dispersions of the atomic densities are shown magnified fivefold. We found that the Cu atoms, which are displaced randomly as the initial condition, move toward the face-centered cubic positions as time passes. Therefore, our phase-field modeling has the ability to restore the equilibrium state of Cu. We then calculated the time evolution of the pressure using the following equation (the derivation is given in Supplemental Material 2\mbox{\cite{PhysRevB.98.024106}}):
\begin{eqnarray}
 P&=&-\Big(\frac{\partial F}{\partial V}\Big)_{T,N}\\ \nonumber
 &=&\frac{N k_{B}T}{V}+\Big \langle-\frac{\partial \phi}{\partial V}\Big \rangle_{\rho_{1},\rho_{2},...,\rho_{N}},
\end{eqnarray}
where $V$ is the volume of the system. Therefore, the total pressure is the expectation value of the pressure $-\partial \phi/ \partial V$ calculated using machine-learning potentials. Figure \mbox{\ref{fig:time}}(b) shows the time evolution of the pressure calculated by the phase-field simulation (red line). For comparison, the pressure calculated by MD is also plotted (black line). We found that the pressures in both the phase-field and MD simulations reach the equilibrium pressure through a similar steep descent behaviour. The equilibrium pressure is 4.86 GPa in the phase-field simulation and 4.76 GPa in the MD simulation. In other words, the pressure obtained by the phase-field simulation is compatible with that obtained by the MD simulation. Thus, we conclude that our modeling successfully reproduces the time evolution of the pressure.

\clearpage

\clearpage
\subsection{Tempereture dependence}

\begin{figure}
  \includegraphics{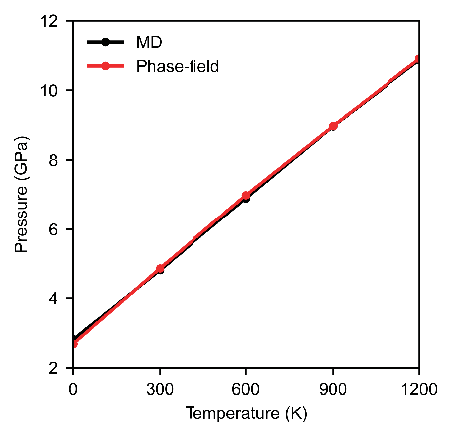}
  \caption{Equilibrium pressures of bulk Cu obtained by phase-field and MD simulations under NVT conditions. The lattice constant is set to 3.615 \mbox{\AA}.}
  \label{fig:nvt}
\end{figure}

\begin{figure}
  \includegraphics{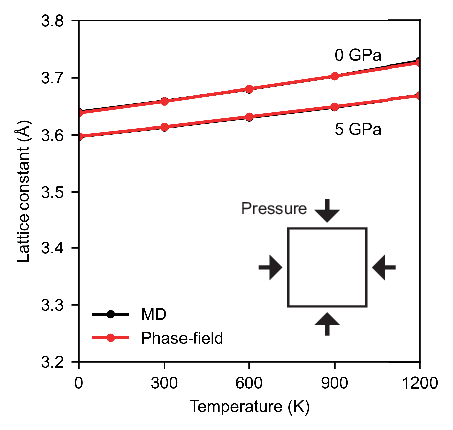}
  \caption{Equilibrium lattice constants of bulk Cu under NPT conditions obtained by phase-field and MD simulations.}
  \label{fig:npt}
\end{figure}

To investigate temperature dependence, we calculated the equilibrium pressure at various temperatures in the NVT ensemble. Figure \mbox{\ref{fig:nvt}}(a) and Table S1 show the relationship between temperature and pressure obtained by phase-field and MD simulations. We found that the pressures from both simulations agree with each other even at 1200 K. Thus, our phase-field modeling can reproduce the temperature dependence of the free energy of atomic systems. To further validate the temperature dependence, we consider the NPT (constant temperature and pressure) ensemble. For this, we change the system volume according to the following equation\mbox{\cite{PhysRevLett.114.155501}}:
\begin{eqnarray}
\frac{dV}{dt}=M(P-P_{0}),
\end{eqnarray}
where $P_{0}$ is the applied external pressure and $M$ is the volume mobility. In other words, we scale the positions and the volume at each time step as follows:
\begin{eqnarray}
\boldsymbol{X_{i}} &\rightarrow& \mu \boldsymbol{X_{i}}, \nonumber \\
V &\rightarrow& \mu^{3} V,
\end{eqnarray}
where $\mu$ is a scale factor, given by
\begin{equation}
\mu=\{1+M' \Delta t (P-P_{0})\}^{\frac{1}{3}},
\end{equation}
where $M'=M/V$ is the mobility per unit volume, which is $M' = 5.0 $ Pa$^{-1}$$\cdot$s$^{-1}$ in this study. Using the above pressure and volume control, we calculated the equilibrium lattice constants by phase-field simulations. Figure \mbox{\ref{fig:npt}} and Table S2 show the equilibrated lattice constants obtained by phase-field and MD simulations with $P_{0}$ = 0 and 5 GPa. We found that the lattice constants calculated by both simulations agree with each other. Therefore, again, the free energy and its derivatives formulated in this study are valid.

\clearpage

\clearpage
\subsection{Atomic-scale thermodynamic distributions around $\Sigma$ 5(310)[001] Cu grain boundary}
\begin{figure}
  \includegraphics{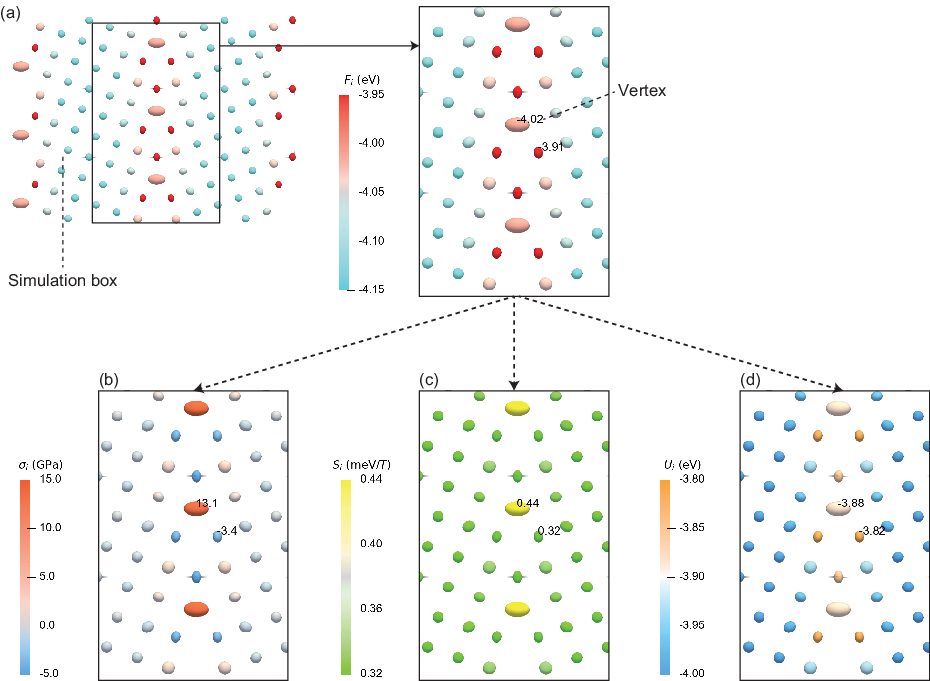}
  \caption{Spatial distributions of (a) free energy, (b) hydrostatic stress, i.e., $(\sigma_{xx}+\sigma_{yy}+\sigma_{zz})/3$, (c) entropy, and (d) potential energy around $\Sigma$5(310)[001] Cu grain boundaries. A simulation box used in the simulations was generated from two rotated bulk Cu crystals with a lattice constant of 3.615 \mbox{\AA}. Then the atomic positions and cell parameters were optimized in the NPT ensemble under ambient conditions ($T=300$ K and $P_{0}=0.1$ MPa). The resulting supercell size is 22.83 $\times$ 5.73 $\times$ 7.28 \mbox{\AA} with 76 atoms.}
  \label{fig:GBdistribution}
\end{figure}

As shown above, the formulated free-energy functional reproduces the free energy of Cu.
We now decompose this free energy into per-atom contributions in order to examine the spatial distribution of local thermodynamic properties.
In a machine-learning interatomic potential, an atomic system is represented as a graph, and any extensive quantity is obtained from the states of the individual nodes. Accordingly, the total free energy $\phi$ is written as follows:
\begin{equation}
\phi=\sum_{i=1}^{N} \Big(\sum_{t=1}^{T} R_{t}(h_{i}^{t})\Big)=\sum_{i=1}^{N}\phi_{i},
\end{equation}
where $h_{i}^{t}$ is the state of node (atom) $i$ after the $t$-th message-passing update, $T$ is the number of such updates, and $R_{t}$ is the corresponding read-out function. The inner sum thus yields the atomic energy $\phi_{i}$ carried by atom $i$. Since energies in universal potentials should be consistent with any systems different from classical potentials, we assume that atomic energies can be regarded as a physical quantitiy like chemical potentials. By substituting this to free energy functional (Eq. \mbox{\ref{eq:freeene}}), their temperature dependencies, that is, an atomic free energy $F_{i}$ can be calculated as follows:
\begin{eqnarray}
\label{eq:atomic_freeene}
F&=& \sum_{i=1}^{N}F_{i}\\ \nonumber
&=&\sum_{i=1}^{N} \Bigg[ \frac{1}{2}k_{B}T \left \{ \ln \left(\frac{\Lambda^{6}}{8\pi^{3} \sigma_{i,x'}^2 \sigma_{i,y'}^2 \sigma_{i,z'}^2} \right)-3 \right\} \\ \nonumber
&+&\int\int...\int{\rho_{1}(\boldsymbol{x}_{1})}{\rho_{2}(\boldsymbol{x}_{2})}...{\rho_{N}(\boldsymbol{x}_{N})}\phi_{i}(\boldsymbol{x}_{1},\boldsymbol{x}_{2},...,\boldsymbol{x}_{N})d\boldsymbol{x}_{1}d\boldsymbol{x}_{2}...d\boldsymbol{x}_{N} \Bigg]. \nonumber
\end{eqnarray}
This representation enables visualization of the spatial distribution of local thermodynamic states with atomic resolution. As an application of this methodology, we visualize local thermodynamic states around a grain boundary in the following. Figure \mbox{\ref{fig:GBdistribution}}(a) shows the free energy distribution around the $\Sigma$5(310)[001] grain boundary of Cu. We find that the free energy locally increases around the grain boundary compared with grain regions. Moreover, even within the grain boundary region, the free energy is spatially inhomogeneous, and the vertices of the structure show relatively smaller values than nearby atoms; their spread reaches $(-4.02)-(-3.91)\approx 0.11$ eV.

To elucidate the origin of the inhomogeneous free-energy distribution, we first computed the atomic stress as follows:
\begin{eqnarray}
\sigma_{i,\alpha \beta}&=&\frac{1}{V_{i}}\bigg(\frac{\partial F_{i}}{\partial \varepsilon_{\alpha \beta}}\bigg)_{T,N} \\ \nonumber
 &=&-\frac{1}{V_{i}} \Big\{k_{B}T+\Big \langle-\frac{\partial \phi_{i}}{\partial \varepsilon_{\alpha \beta}}\Big \rangle_{\rho_{1},\rho_{2},...,\rho_{N}} \Big\},
\end{eqnarray}
where $V_{i}$ is an atomic volume defined via Voronoi tessellation in this study\mbox{\cite{10.1063/1.3215722,PhysRevE.74.021306}}, and $-\frac{\partial \phi_{i}}{\partial \varepsilon_{\alpha \beta}}$ is the atomic virial, evaluated on the interaction graph as follows:
\begin{eqnarray}
-\frac{\partial \phi_{i}}{\partial \varepsilon_{\alpha \beta}}=\sum_{j\in N(i)}\frac{1}{2}[\mathbf{R}_{ij}\otimes \mathbf{F}_{ij}+(\mathbf{R}_{ij}\otimes \mathbf{F}_{ij})^{T}]_{\alpha \beta},
\end{eqnarray}
where $\mathbf{R}_{ij}$ is the interatomic vector and $\mathbf{F}_{ij}=-\partial \phi/\partial \mathbf{R}_{ij}$ is the force between atom $i$ and its neighbor $j \in N(i)$. Figure \mbox{\ref{fig:GBdistribution}}(b) shows the atomic-stress distribution around the grain boundary. We find that the vertices of the triangular motif experience tensile stress, whereas other sites are under compression or comparatively weak stress. This behavior is primarily governed by local interatomic distances: larger separations tend to generate tension, while shorter separations produce compression.


To further understand the origin of the inhomogeneous free-energy distribution, we computed the atomic entropy as follows:
\begin{eqnarray}
 S_{i}&=&-\Bigr(\frac{\partial F_{i}}{\partial T}\Bigl)_{V,N} \\ \nonumber
 &=&-\frac{1}{2}k_{B}\left\{\ln \left(\frac{\Lambda^{6}}{8\pi^{3} \sigma_{i,x'}^2 \sigma_{i,y'}^2 \sigma_{i,z'}^2} \right)-6\right\}. \nonumber
\end{eqnarray}
Figure \mbox{\ref{fig:GBdistribution}}(c) shows the atomic-entropy distribution around a grain boundary. We found that the entropy is locally enhanced at the vertices of the triangular motif, coinciding with the sites of highest tensile stress. These observations indicate an asymmetric influence of stress on entropy at the atomic scale: under tensile stress, additional free space permits larger thermal vibrations, increasing 
$S_i$ because entropy reflects the width of the atomic vibration; under compressive stress, the available space is reduced, and $S_i$ cannot increase as under tension. Consequently, the entropic contribution to the free energy, $-TS_i$, becomes spatially inhomogeneous around the grain boundary in accordance with the local stress state.

On the other hand, from the calculated entropies we can obtain the per-atom potential energy as follows:
\begin{eqnarray}
 U_{i}&=&F_{i}+TS_{i}
\end{eqnarray}
Figure \mbox{\ref{fig:GBdistribution}}(d) shows the potential-energy distribution around a grain boundary. We find that the potential energy is also spatially inhomogeneous, similarly to the free energy, but its spread is smaller, that is, about 0.06 eV. Thus, the inhomogeneous free-energy landscape arises from the combined effects of inhomogeneous entropy and potential energy around the boundary. Note that the potential-energy distribution does not necessarily correspond to the stress distribution; for example, atoms under the highest tensile stress do not exhibit the highest potential energy. This is because the potential energy (i.e., atomic stability) is determined by both elastic and chemical contributions governed by local bonding, $U_{i}=U^{\rm{elastic}}_{i}+U^{\rm{chemical}}_{i}$.

The above observations suggest that the free-energy distribution around a grain boundary is formed by two steps: (1) The potential energy is spatially inhomogeneous due to inhomogeneous atomic configurations near the boundary, producing the first inhomogeneity in the free-energy distribution. (2) At the same time, the inhomogeneous configurations also introduce an inhomogeneous stress distribution, that is, some atoms undergo tension while others experience compression. As a result, entropy generated at finite temperature is also spatially inhomogeneous, which further introduces spatial variations in free energy. As these two inhomogeneities accumulate in the case of the $\Sigma$5(310)[001] Cu grain boundary, the free-energy differences among atoms can reach 0.1 eV or more. (3) Significance of the decomposition: Free energy around grain boundaries are commonly used to assess their stability via quantities such as interfacial and segregation energies. In contrast, our methodology not only provides the free energy but also its spatial distribution. Consequently, the proposed framework yields not only stability information for a given grain boundary but also mechanistic insight into why it is stabilized, by visualizing free energy, stress, and entropy at atomic resolution.

\clearpage
\section{Conclusion}
In summary, we have developed atomic-scale phase-field modeling with universal machine-learning potentials. Our modeling successfully reproduced the pressure and lattice constant of Cu under the NVT and NPT ensembles. Furthermore, this modeling enables us to decompose the free energy into contributions from each atom, providing local distributions of free energy, stress, entropy, and potential energy around the $\Sigma5(310)[001]$ Cu grain boundaries. Our modeling extends phase-field modeling, that is, local thermodynamic theory to the ultra-small scale, enabling the characterization of material properties with atomic resolution.

\begin{acknowledgement}
This work was supported by JSPS KAKENHI grant numbers 25K23433.

\end{acknowledgement}

\begin{suppinfo}
The Supporting Information is available free of charge at
\begin{itemize}
  \item PDF: Derivative of the Free Energy with Respect to the Gaussian Width, Derivation of the pressure formula, Derivation of atomic virials in graph-based models, the equilibrium pressure under the NVT condition (Table S1), the equilibrium lattice constant under the NPT condition (Table S2), and a movie for atom dynamics.
\end{itemize}

\end{suppinfo}\textbf{}

\bibliography{manuscript}

\end{document}